\documentstyle[prl, aps, twocolumn, epsfig]{revtex}

\newcommand\be{\begin{equation}}
\newcommand\ee{\end{equation}}
\newcommand\bd{\begin{displaymath}}
\newcommand\ed{\end{displaymath}}
\newcommand\bea{\begin{eqnarray}}
\newcommand\eea{\end{eqnarray}}
\newcommand\nn{\nonumber}
\newcommand\lt{\left}
\newcommand\rt{\right}

\begin{document}

\twocolumn[

\hsize\textwidth\columnwidth\hsize\csname @twocolumnfalse\endcsname

\preprint{Fermilab Preprint: Pub-98/270-A}

\draft

\title{Higher-Order Methods for Quantum Simulations}

\author{A. T. Sornborger and E. D. Stewart}

\address{NASA/Fermilab Astrophysics Group, Fermi National Accelerator
Laboratory, Box 500, Batavia, IL 60510-0500, USA}

\date{\small September 3, 1998}

\maketitle

\begin{abstract}
To efficiently implement many-particle quantum simulations
on quantum computers we develop and present methods
for inverting the Campbell-Baker-Hausdorff lemma to 3rd
and 4th order in the commutator. That is, we reexpress
$ \exp \lt\{ -i \lt( H_1 + H_2 + \dots \rt) \Delta t \rt\} $
as a product of factors $ \exp \lt( -i H_1 \Delta t \rt) $,
$ \exp \lt( -i H_2 \Delta t \rt)$, $\dots$ which is accurate
to 3rd or 4th order in $\Delta t$.
\end{abstract}


]

Quantum computers have generated much interest recently, largely due
to the result by Shor \cite{shor} that they can factor integers in
an amount of time that grows polynomially with the size of the
integer. This can be compared to factorization on a classical
computer, where the time it takes to factor a number grows
exponentially with the input size. In addition to Shor's factorization
algorithm, simulations of quantum systems have also been shown to be
possible in polynomial time \cite{simulate}. Indeed, this was the
first area for which it was proposed that quantum computers could
fundamentally be more powerful (i.e. much faster) than classical
computers \cite{feynman}.

From a theoretical standpoint, a quantum computer is a quantum system
with a $2^n$-dimensional Hilbert space. Pairs of states in the system
are defined to be `qubits'. The canonical example of such a system is
a set of $n$ spins. Each spin consists of two states, so each spin can
represent a qubit and the Hilbert space of the system is
$2^n$-dimensional. The equivalent of a logical gate on a classical
computer is an operator acting on a set of qubits on a quantum
computer.

This letter focuses on a problem which concerns simulational issues in
quantum computation. A simulation of a quantum mechanical system on a
quantum computer consists of applying an operator $\exp(-i H t)$ on a
set of qubits, where $H$, the Hamiltonian of the system of interest,
is suitably encoded (and discretized) to act on the set of qubits. For
many-particle systems $H$ is a sum of terms. For instance, the Hubbard
model Hamiltonian, used in the study of high-$T_c$ superconductivity,
can be written \cite{abramslloyd} as the sum 
\be
H = \sum_{i=1}^m V_0 n_{i\uparrow} n_{i\downarrow} + \sum_{\langle i,
j \rangle \sigma} t_0 c_{i \sigma}^* c_{j \sigma} \label{ham}
\ee
where $V_0$ is the strength of the potential, and $n_{i\sigma}$ is the
operator for the number of fermions of spin $\sigma$ at site $i$. In
the second (kinetic energy) term, the sum $\langle i, j \rangle$
indicates all neighboring pairs of sites, $t_0$ is the strength of the
``hopping'', and $c_{i\sigma}$, $c_{i\sigma}^*$ are annihilation and
creation operators, respectively, of a fermion at site $i$ and spin
$\sigma$. This model gives an example in which a full simulation on a
classical computer is impossible due to the exponential increase in
the size of the Hilbert space of the quantum system with the number of
lattice sites.

The canonical quantum computer cannot act on all spins at
once \cite{qc}. Therefore, it becomes necessary to find ways of
approximating the evolution operator, which is the exponential of
a sum of operators (with a Hamiltonian such as that in
Eq.~(\ref{ham})) as a product of operators each acting on a subspace
of the Hilbert space. To second order, for instance, we could use the
approximation
\bea
\lefteqn{
\lt( e^{-i H_1 \Delta t} e^{-i H_2 \Delta t} \ldots
 e^{-i H_N \Delta t} \rt)
} \nn \\
\lefteqn{
\lt( e^{-i H_N \Delta t} \ldots e^{-i H_2 \Delta t}
 e^{-i H_1 \Delta t} \rt)
} \nn \\
& = &
e^{ -i 2 \lt( H_1 + H_2 + \ldots + H_N \rt) \Delta t
 + {\cal O} \lt[ \lt( \Delta t \rt)^3 \rt] }
\eea
where the $e^{-i H_n \Delta t}$ act on a subspace of the Hilbert
space.

To find higher order approximation methods, we want to reexpress
$\exp\lt(\sum_{n=1}^N A_n\rt)$ as a product of individual
$\exp\lt(A_n\rt)$'s. In order to do this, we must invert the
Campbell-Baker-Hausdorff formula. To 5th order, the
Campbell-Baker-Hausdorff formula is
\bea
\lefteqn{
\exp \lt( a A_1 \rt) \exp \lt( a A_2 \rt)
= } \nn \\ &&
\exp \lt[ a \lt( A_1 + A_2 \rt) + \frac{1}{2} a^2 A_{12}
+ \frac{1}{12} a^3 \lt( A_{112} + A_{221} \rt)
\rt. \nn \\ && \lt. \mbox{}
+ \frac{1}{24} a^4 A_{1221}
- \frac{1}{720} a^5
\lt( A_{11112} - 2 A_{21112} - 6 A_{11221}
\rt. \rt. \nn \\ && \lt. \lt. \mbox{}
- 6 A_{22112} - 2 A_{12221} + A_{22221} \rt)
+ {\cal O} \lt( a^6 \rt) \rt]
\label{CH} \eea
where
\be \label{A}
A_{kl \ldots mn} \equiv
\lt[ A_k , \lt[ A_l , \ldots \lt[ A_m , A_n \rt] \ldots \rt] \rt]
\ee

As a strategy for finding approximation methods, we pick a fundamental
ordering of the product of exponentials with parameters allowing for
transposes of the entire product as well as raising all the
exponentials in the fundamental unit to the same power. By iterating
the Campbell-Baker-Hausdorff formula, we can get an expression for
this fundamental unit in terms of a single exponential
\be \label{unit}
\lt( e^{a A_1} e^{a A_2} \ldots e^{a A_N} \rt)^\alpha
= \exp \sum_{p=1}^\infty \alpha a^p B_N^p
\ee
which defines the $B_N^p$ in terms of the $A_n$. Here, $p$ is an
exponent on $a$, and a label on the matrices $B_N^p$.

Now combine a succession $i = 1, \dots, I$ of fundamental units with
parameters $a_i$ and $\alpha_i$. Again iterating
Campbell-Baker-Hausdorff gives
\bea \label{BX}
\exp \lt( \sum_{p=1}^\infty \alpha_1 a_1^p B_N^p \rt)
\ldots
\exp \lt( \sum_{p=1}^\infty \alpha_I a_I^p B_N^p \rt) \nn \\
= \exp \lt( \sum_X \sigma_I^X B_N^X \rt)
\eea
The $B_N^X$ are generated from the $B_N^p$ by commutation. $X$
represents a label ${pq \ldots rs}$ where
\be
B_N^{pq \ldots rs} \equiv
\lt[ B_N^p , \lt[ B_N^q , \ldots \lt[ B_N^r , B_N^s \rt] \ldots \rt] \rt]
\ee
$B_N^{pq \ldots rs}$ is of order $ p + q + \ldots + r + s $.
Up to 5th order we can take
\be
X \in \lt\{ 1; 2; 3, 12; 4, 13, 112; 5, 14, 23, 113, 221, 1112 \rt\}
\ee
These $B_N^X$ span the space of the $B_N^p$'s and their commutators to
5th order and for $N \geq 2$ they are independent. The $\sigma_I^X$
are defined in terms of $\alpha_i$ and $a_i$ by Eq.~(\ref{BX}). Here
again, the $X$'s are labels. After some calculation, the
Campbell-Baker-Hausdorff formula, Eq.~(\ref{CH}), gives the
equations
\be \label{sigp}
\sigma_I^p = \sum_{i=1}^I \alpha_i {a_i}^p
\ee
for $ p = 1, \ldots, 5 $,
\be
\sigma_I^{pq} = - \frac{1}{2} \sigma_I^p \sigma_I^q 
+ \frac{1}{2} \sum_{i=1}^I {a_i}^{q-p}
\lt[ \lt( \sigma_i^p \rt)^2 - \lt( \sigma_{i-1}^p \rt)^2 \rt]
\ee
for $pq = 12, 13, 14, 23$,
\bea
\lefteqn{
\sigma_I^{ppq} = - \frac{1}{2} \sigma_I^p \sigma_I^{pq}
- \frac{1}{6} \lt( \sigma_I^p \rt)^2 \sigma_I^q
} \nn \\ &&
+ \frac{1}{6} \sum_{i=1}^I {a_i}^{q-p}
\lt[ \lt( \sigma_i^p \rt)^3 - \lt( \sigma_{i-1}^p \rt)^3 \rt]
\label{sigppq} \eea
for $ppq = 112, 113, 221$, where $\sigma_I^{21} \equiv -\sigma_I^{12}$
\bea
\sigma_I^{1112} = - \frac{1}{2} \sigma_I^1 \sigma_I^{112}
- \frac{1}{3} \lt( \sigma_I^1 \rt)^2 \sigma_I^{12}
- \frac{1}{24} \lt( \sigma_I^1 \rt)^3 \sigma_I^2 \nn \\
+ \frac{1}{24} \sum_{i=1}^I a_i
\lt[ \lt( \sigma_i^1 \rt)^4 - \lt( \sigma_{i-1}^1 \rt)^4 \rt]
\eea

For approximations to $\exp\lt(\sum_{n=1}^N A_n\rt)$, we require all
$\sigma_I^X = 0$ except for $\sigma_I^1$ which is the coefficient of
$B_N^1 = \sum_{n=1}^N A_n$, and which should be greater than zero.

An interesting feature of 3rd order methods is that they require
at least one inverse, \mbox{i.e.} they require backward time evolution
during part of the method.\footnote{After this work was completed, we
became aware that this point had also been noted in
\cite{otherwork}.} This follows immediately from Eq.~(\ref{sigp}) with
$p = 3$. It can also be proved using Eq.~(\ref{sigp}) with $p = 3$ and
$p = 4$ that 4th order methods must have at least two inverses.

Our basic method to solve Eqs.~(\ref{sigp}-\ref{sigppq}) is to pick
values of $\alpha_i$ and $a_i$ and see if they satisfy the
equations. To do this we must restrict the number of fundamental units
by fixing $I$. We also take the $\alpha_i$'s to be $\pm 1$ and
restrict the range of the $a_i$'s.

We start with Eq.~(\ref{sigp}), since, in this equation, order with
respect to $i$ does not matter. So, for a given set of values, we need
to consider only one permutation, not all permutations of the
values. This greatly reduces the size of the search.

Furthermore, we start by considering $p = 1$ and $3$, since it is only
the sign of $\alpha_i a_i$ that matters in these equations. This
means we can consider only the sign of the combination $\alpha_i
a_i$, and not the signs of $\alpha_i$ and $a_i$ individually. This
reduces the search further. These equations are particularly
restrictive for the case of few inverses.

After solving the $p = 1$ and $3$ equations, we introduce separate
signs for the $\alpha_i$'s and $a_i$'s and solve the equation with $p
= 2$, and $p = 4$ for the 4th order case.

Finally, into the restricted set of solutions to Eq.~(\ref{sigp}) we
introduce permutations of the $\alpha_i$'s and $a_i$'s with respect to
the index $i$ and solve Eqs.~(\ref{sigp}-\ref{sigppq}).

We find a larger number of solutions than we can easily present. We
want to present solutions which are in some sense optimal. To do this,
we consider the form of the operator resulting from a given method
\bea
\prod_{j=1}^I \lt( e^{-i a_j A_1 \,\Delta t} \, e^{-i a_j A_2 \, \Delta t}
\ldots e^{-i a_j A_N \, \Delta t} \rt)^{\alpha_j} \nn \\
= \exp \lt[ -i \sigma_I^1 \sum_{n=1}^N A_n \,\Delta t
 + r (-i \,\Delta t)^{o+1} \rt]
\eea
where $\Delta t \ll 1$ is a time step, $o$ is the order of the method,
and
\be\label{r}
r = \sum_X \sigma_I^X B_N^X
\ee
where $X \in \{4, 13, 112\}$ for a 3rd
order method and $X \in \{5, 14, 23, 113, 221, 1112\}$ for a 4th order
method.

$r$ is an error which takes values in the vector space of the
commutators for which we do not have a metric. Therefore, we make an
{\it ad hoc\/} choice of basis to be discussed elsewhere. This allows
us to replace $r$ by a single real scalar $R$. The error from the
method can then be taken to be
\be\label{E}
E = n R \, \Delta t^{o+1}
\ee
where $n$ is the number of times we apply the
approximate method.

If the physical time we want to simulate is $T_p$, then
\be\label{TP}
T_p = n D \Delta t
\ee
where $D \equiv \sigma_I^1$ is given by the method.

The computer time it takes for a given simulation can be written
\be 
T_c = n I N t_g + n L N t_s
\ee
where $I$ is the number of fundamental units in the method and $N$ is
the number of terms in a unit, $t_g$ is the time it takes to make a
gate change,
\be
L \equiv \sum_{i = 1}^I |a_i|
\ee
so that $L N$ is the total time the gates are applied for in the
method. The time an individual gate is applied for will be $t_s = b \,
\Delta t$, where $b$ is a proportionality constant dictated by the
actual couplings in the quantum computer hardware.

Using Eqs.~(\ref{E}) and~(\ref{TP}), the computer time can be
rewritten 
\be
T_c = \lt( \frac{T_p^{o+1}}{E} \rt)^{\frac{1}{o}}
\lt( \frac{G}{D} \rt) \lt( \frac{R}{D} \rt)^{\frac{1}{o}} t_g
+ \frac{LbT_p}{D}
\ee

There are two possible limits to this equation. One is that the
computer time is dominated by gate switching. In this case, we want
the factor
\be
Z = \lt( \frac{G}{D} \rt) \lt( \frac{R}{D} \rt)^{\frac{1}{o}}
\ee
to be small. The second limit is when the computer time is dominated
by the time during which the gates are applied. Here, the ratio $L/D$
should be small, and to make the error small, we want $(R/D)(\Delta
t)^o$ small. However, if $\Delta t$ can be made very small (from the
hardware point of view), then making the error small forces the
computer time to be dominated by gate switching. If there is a limit
to $\Delta t$, and it is reached before the computer time is gate
switching dominated, then the computer time may still be dominated by
gate application and we want $L/D$ and $R/D$ small. We also prefer to
have concise methods.

The 3rd order method that we have selected given the above criteria is
\be
(1)^T(1)(1)(1)(1)^T(-2)^T(1)(1)(1)
\ee
and the 4th order method is 
\bea
(1)^T(1)(1)^T(-2)(1)^T(1)^T(1)^T(1)^T(1) \nn \\
(1)^T(1)(1)(1)(1)(-2)^T(1)(1)^T(1)
\eea
where $(x)$ denotes
\be
(e^{xA_1} e^{xA_2} \ldots e^{xA_N})
\ee
and $(x)^T$ denotes
\be
(e^{xA_N} \ldots e^{xA_2} e^{xA_1})
\ee

To illustrate our methods, we have applied first, second, third and
fourth order methods to the exactly soluble operator
\be
e^{ -i \, \Delta t \lt( \sigma_x + \sigma_y + \sigma_z \rt) } =
\lt( \begin{array}{cc}
C - \frac{i}{\sqrt{3}} S & -\frac{1+i}{\sqrt{3}} S \\
\frac{1-i}{\sqrt{3}} S & C + \frac{i}{\sqrt{3}} S
\end{array} \rt)
\ee
where $C \equiv \cos \lt( \sqrt{3} \, \Delta t \rt)$ and
$S \equiv \sin \lt( \sqrt{3} \, \Delta t \rt)$.

As a measure of the error, we took the differences $\Delta\sigma_x$,
$\Delta\sigma_y$ and $\Delta\sigma_z$ between the $\sigma_x$,
$\sigma_y$ and $\sigma_z$ components of the exact solution and those
of the results of our methods. We then calculated the error
\be
E = \sqrt{ \lt( \Delta\sigma_x \rt)^2 + \lt( \Delta\sigma_y \rt)^2
 + \lt( \Delta\sigma_z \rt)^2 }
\ee

\begin{figure}
\psfig{figure=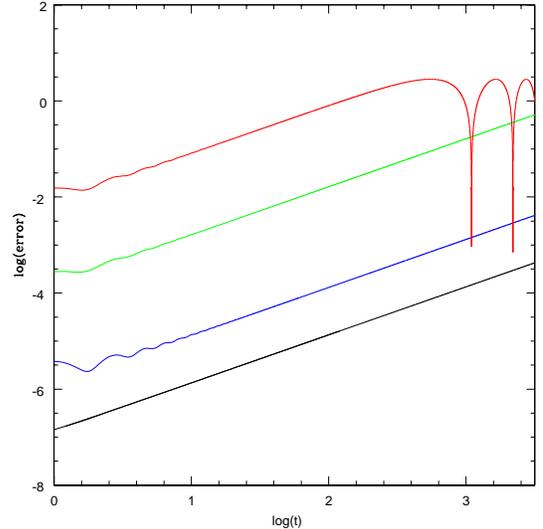,height=3.0in,width=3.0in}
\caption[caption]{A measurement of the accuracy of our
results. Plotted is the $\log(error)$ of (from top to bottom) first,
second, third and fourth order approximation methods as a function of
$\log(time)$. Note that for the fourth order method, the error never
grows larger than $10^{-3}$.
\label{singlespin}}
\end{figure}

In Fig.~(\ref{singlespin}), we plot the logarithm of the error as a
function of the logarithm of the time that the system was evolved
for. The first order method results are uppermost and higher order
results lie underneath each other with fourth order results being the
lowermost plotted. $\Delta t = 0.01$ for all methods.

Notice that the first order error oscillates once it reaches order
$1$. The rest of the errors remain small throughout the simulation,
with the fourth order error remaining below $10^{-3}$ for the entire
evolution.

The error for all methods goes as $n R \lt( \Delta t \rt)^{o+1}$,
where $n$ is the number of times the method has been applied.
Therefore,
$ \log E = \log n + \log \lt[ R \lt( \Delta t \rt)^{o + 1} \rt] $.
For $\Delta t = 0.01$, this makes the $y$-intercept decrease roughly
by order $-2$ as the order of the method increases. Since the time
evolved is proportional to $n$, the slope of the errors is $1$ for
all methods.

As an example of how useful our approximations can be, let us consider
a case in which we want to apply an approximation method for time $T =
1$ with total error $E = 10^{-4}$. For a first order method, this
means that we require about $5000$ applications of the method. For
second order, we require about $30$ applications. For our third order
method, we need $2$ applications. And for our fourth order method, we
need less than $1$ application of the method. This results in a
reduction of orders of magnitude in the computational cost of a given
simulation.

\section*{Acknowledgements}

This work was supported by the DOE and the NASA grant NAG 5-7092 at
Fermilab. We would like to thank Tasso Kaper for bringing references
\cite{otherwork} to our attention.

\end{document}